# How additive manufacturing can boost the bioactivity of baked functional foods

Sara M. Oliveira[1], Alice Gruppi[1,2], Marta V. Vieira[1], Gabriela M. Souza[3], António A. Vicente[3], José A.C. Teixeira[3], Pablo Fuciños[1], Giorgia Spigno[2], Lorenzo M. Pastrana[1,*]

[1]International Iberian Nanotechnology Laboratory
Av. Mestre José Veiga s/n
4715-330 Braga - Portugal
*E-mail: lorenzo.pastrana@inl.int

[2]Catholic University of the Sacred Heart
Department for Sustainable Food Process
Via Emilia Parmense, 84-29122 Piacenza, Italy

[3]CEB – Centre of Biological Engineering, University of Minho,
Campus de Gualtar, 4710-057 Braga, Portugal

*corresponding author: lorenzo.pastrana@inl.int

**Abstract**

The antioxidant activity of baked foods is of utmost interest when envisioning enhancing their health benefits. Incorporating functional ingredients is challenging since their bioactivity naturally declines during baking. In this study, 3D food printing and design of experiments are employed to clarify how the antioxidant activity of cookies enriched with encapsulated polyphenols can be maximized. A synergistic effect between encapsulation, time, temperature, number of layers, and infill of the printed cookies was observed on the moisture and antioxidant activity. Four-layer cookies with 30 % infill provided the highest bioactivity and phenolic content if baked for 10 min and at 180 °C. The bioacitivity and total phenolic content improved by 115 % and 173 %, respectively, comparing to free extract cookies.

Moreover, the proper combination of the design and baking variables allowed to vary the bioactivity of cooked cookies (moisture 3-5 %) between 300 to 700 μmolTR/gdry. The additive manufacture of foods with interconnected pores could accelerate baking and browning, or reduce thermal degradation. This represents a potential approach to enhance the functional and healthy properties of cookies or other thermal treated bioactive food products.

## 1. Introduction

The recognized tie between diet and health, along with the ever-increasing life expectancy, have promoted the awareness of consumers towards food and its potential health benefits (Bigliardi and Galati, 2013; Day et al., 2009; Karelakis et al., 2020). Food may provide benefits over physiological functions, preventing and treating diseases, such as obesity, cancer, or diabetes (Battino et al., 2019; Vieira da Silva et al., 2016). Delivering high quality functionalized and health-driven products through sustainable food processes is one of the most significant challenges food research is currently facing (Wang and Boh, 2012).

The incorporation of biologically-active compounds into food is directed to ingredients with well-accepted health benefits, such as carotenoids, polyphenols, vitamins, and some proteins (Augustin and Sanguansri, 2015; Chew et al., 2019). Polyphenols, in particular, are ubiquitous plant secondary metabolites, with several pharmacological effects, mostly led by their remarkable antioxidant capacity (Costa et al., 2017; Daglia, 2012; Squillaro et al., 2018). The efficient enrichment of food matrices is challenging, as the polyphenols are sensitive substances undergoing numerous reactions in the course of food processing (Souza et al., 2017). Cooking may cause complex physicochemical changes in phenolic compounds, including release from bound forms, degradation, polymerization, and oxidation (Mcclements et al., 2009; Palermo et al., 2014).

Among the food functionalization strategies, nano/micro-encapsulation is a commonly adopted one. Encapsulation aims at protecting labile functional ingredients against external conditions, reducing undesired tastes, and improving their bioavailability and physicochemical characteristics (Akbari-Alavijeh et al., 2020; Assadpour and Jafari, 2019; Shtay et al., 2019). Alongside is the increasing pursuit of sustainable food personalization focusing on different sensorial and nutritional needs, e.g., children *vs*. elderly (Rodgers, 2016).

From this perspective, 3D food printing, i.e., additive manufacturing of foods, emerges as an exciting technology to customize sensorial attributes (e.g., shape, texture, color) and nutrition to each individual (Dankar et al., 2018; Lipton et al., 2015). Although significant progress in automation, scaling-up, and genuine personalization of the process are still needed, the application on some foods, such as mashed potatoes, doughs, chocolate, cheese, and meat, have already been studied (Godoi et al., 2016; Kumar et al., 2020; Le-Bail et al., 2020).

Cookies are a snack widely consumed across all ages. Their functional enrichment is complicated by the baking temperature and time required to reach their distinctive moisture (< 5% (Charissou et al., 2007)) and browning. Complex reactions involving protein denaturation, fat-melting, starch granule-loss, and surface browning by caramelization and Maillard reactions develop its brown color and crumbly texture (Ameur et al., 2006; Chevallier et al., 2000; Purlis, 2010). The heat and (vapor and water) mass transfer establish the moisture content and its distribution (Demirkol et al., 2006; Thorvaldsson and Janestad, 1999). Mass diffusion starts outwards to the food surface. Its transfer coefficient is determined by the thermo-physical properties of the medium and the food (i.e., shape, size, pore, pore connectivity, surface roughness, and temperature); whereas the heat transfer depends on the conductivity properties (Ahmad et al., 2001; Demirkol et al., 2006; Gueven and Hicsasmaz, 2013).

The usual reduction of bioactivity of enriched cookies observed during baking mainly results from the thermal degradation of the functional ingredients (Nayak et al., 2015). The kinetics of the thermal degradation of phenolic compounds depends on specific properties of the food matrix, such as the presence of reducing sugars (Debicki-Pospisil et al., 1983), sucrose (Nikkhah et al., 2007), Maillard reaction products (Namiki, 1988), and ascorbic acid (De Rosso and Mercadante, 2007), in addition to time and temperature (Charissou et al., 2007). The Maillard reaction is a non-enzymatic browning process that generates antioxidant molecules (Virág et al., 2013) and generally initiates when water activity reduces to 0.4-0.7, and the temperature is above 105-120 °C (Purlis, 2010). Therefore, the reaction can contribute significantly to the bioactivity of the cookies. The optimal baking conditions (i.e., time, temperature) rely on how fast the caramelization and Maillard reaction initiate, which provides an opportunity window for improvements. The Maillard reaction occurs when amino acids and reducing sugars are heated, whereas the caramelization relates to carbohydrates reactions (Purlis, 2010). In both cases, the initiation requires high temperature and low water activity at the surface.

The geometric features, such as thickness, pore shape, surface area/volume ratio, porosity, and pore size, are relevant for the convective baking process (Zhang et al., 2018). Optimizing the interactions and effects between those properties might accelerate the reactions without compromising the expected color, aroma, and moisture; while enhancing the antioxidant activity. 3D food printing is a technology that enables us to control such properties in a reproducible manner. In addition to the external shape, it specifies other parameters, such as the number of layers, layer thickness, extrusion diameter, and internal infill and pattern. As a matter of fact, a recent study (Zhang et al., 2018) reported a 2-log fold increase of probiotic activity from $10^4$ CFU/g to $10^6$ CFU/g of the initial $10^9$ CFU/g on baked and porous 3D cereal-based food through low-temperature baking until low moisture was achieved. However, particularly for cookies, the requirements go further, and simultaneous browning, low moisture, and high antioxidant activity are desired. Clearly understanding whether and how the design of 3D printed enriched dough could boost the functionality is a question yet to be answered.

Accordingly, this study centered on modeling the impact of baking (temperature, time) and printing parameters (infill, number of layers) on the bioactivity, color, shape fidelity, and moisture of cookies enriched with microencapsulated polyphenols extracted from grape skins. Furthermore, the study discusses how low moisture cookies with different phenolic content and bioactivity can be obtained from a single formulation.

## 2. Materials and Methods

*2.1. Polyphenols extraction by Ohmic heating*

Fresh grape pomace of Croatina variety was kindly collected after vinification and provided by the winery Mossi 1558 (Ziano Piacentino, Piacenza, Italy) during the 2018 vintage. The pomace was dried at 60 °C until residual moisture content was below 10 % to preserve the antioxidant compounds. Then, dried skins were manually separated and ground into a particle size below 1 mm. Ethanol-based extractions were performed according to the methodology previously described by Ferreira-Santos (Ferreira-Santos et al., 2019) in a glass reactor (400 mL, 150 mm height, and 100 mm diameter), with a double-walled water-jacketed using two electrodes (stainless steel, area of 385 mm$^2$) kept at a 90 mm of distance. The extractions were executed with a total volume of 300 mL by experiment, using a 1:10 solid/liquid ratio of dried powder grape skins and distilled water at 60 °C for 60 min. The solutions presented an average conductivity of 2.5 ± 0.22 mS/cm. The experiments were carried out in triplicate, and the final extract was filtered twice with filter paper under vacuum to eliminate insoluble residues. The grape extract was divided into two lots. The first lot was freeze-dried (Freeze Dryer Lyoquest -55ºC Plus Eco, Telstar) for 3 days, and the obtained powder was included in the dough and is referred to as free extract. The second lot was mixed with maltodextrin and spray-dried to get an encapsulating matrix further added into the dough for baking and is referred to as encapsulated extract.

*2.2. Grape extract encapsulation with Nano Spray-dryer*

The drying process was performed in a laboratory-scale Nano Spray-dryer (Büchi Nano Spray Dryer B-90), with a large diameter nozzle, which corresponds to a range of particles size between

10 and 60 µm. The following operating conditions were adopted: 20 % feeding rate with 12 % w/v total solids concentration, drying gas flow rate (compressed air) of 100 L/min; 75 °C inlet air temperature, and 30 % spray rate. Maltodextrin (DE 13-17, Sigma-Aldrich) was used as a carrier at a proportion of 1:1 maltodextrin dry weight: extract total solids dry weight, as performed by Lavelli *and co-workers* (Lavelli et al., 2016).

## 2.3. Scanning Electron Microscopy

The morphological characterization of the grape extract particles was performed by Scanning Electron Microscopy (SEM), using a Quanta FEG 650 (FEI, USA). Dry samples were fixed on aluminum holders covered by conductive double-sided adhesive carbon tape, followed by a gold coating. Samples were analyzed using an accelerating voltage of 5 kV under vacuum conditions and a magnification of 15,000x. The average particle size was determined by image-based analysis using ImageJ (Image J 1.52i).

## 2.4. Cookie dough preparation

The cookie dough was prepared according to Kim and co-workers (Kim et al., 2019) with the following ingredients: 39.5 % wheat fine flour (type 00, Barilla, Italy), 25 % unsalted butter (LACTOGAL Produtos Alimentares, Porto – Portugal), 22 % powdered sugar (RAR - REFINARIAS DE AÇÚCAR REUNIDAS, Porto – Portugal), 13 % whole milk (LACTOGAL Produtos Alimentares, Porto - Portugal), and 0.5 % xanthan gum (Guzmán, Mercabarna – Spain). For enriched cookie doughs, the grape extract, either free or encapsulated, was added by reducing the sugar percentage correspondently. Briefly, the dried ingredients were mixed, and the liquids were slowly added in a food processor mixing at low speed for 5 min.

The prepared doughs were: blank dough (original recipe above), dough incorporated with free dried extract (19.5 % powdered sugar and 2.5 % free extract), and with encapsulated extract (17 % powdered sugar and 5 % encapsulated extract). The doughs were sealed-stored in the fridge at 4 °C and used after 48 h to allow full hydration of the xanthan gum.

*2.5. Slicing and 3D printing*

A cylinder (27.6 mm diameter, 6.72 mm height) was designed with AutoCAD (v. 2013) and exported as an STL file. The G-code was generated using Simplyfy3D (4.1.2). The model was sliced for a nozzle diameter of 1.6 mm, layer height of 1.12 mm, 1 perimeter, Infill between 30 to 100 %, flow multiplier of 1.12, number of layers between 4 to 6, and a printing speed of 10 mm/s. The G-code was uploaded to a Focus 3D Food Printer (byFlow, the Netherlands). The dough was printed at room temperature and the initial nozzle height was set to zero.

*2.6. Porosity*

The porosity of the sliced models was determined with the following equation:

$$\mathbf{Porosity} = \frac{V_T - V_{sliced\ model}}{V_T} \cdot \mathbf{100}\ (\%)\ \text{(Equation 1)}$$

The total volume of the external sliced shape, total volume (VT) was calculated as:

$$V_T = \pi \cdot radius^2 \cdot N^{\circ}\ Layers \cdot Layer\ thickness\ (cm^3)\ \text{(Equation 2)}$$

The volume of the sliced model, equivalent to the printed volume, was estimated by reverting the plastic volume specified in the G-code scripted by the Simplyfy3D. This volume is calculated considering the density of the plastic filament and the weight to be extruded. The volume should be further rectified with the flow multiplier adjustment used, resulting in the following equation:

$$V_{sliced\ model} = \frac{plastic\ weight}{\rho_{plastic}} \cdot \frac{1}{Multiplier}\ (cm^3)\ \text{(Equation 3)}$$

*2.7. Estimation of surface-to-volume ratio*

The cross-section assumed for the printed filaments was ellipsoidal, whose major axis is considered half the nozzle diameter, and the minor axis is half of the layer thickness. The total amount of extruded filament can be consulted in the G-Code file of each model.

The surface area of the total extruded filament is given by:

$$Surface\ area = Lenght_{Filam} \times Elippse_{perimeter}\ \text{(Equation 4)}$$

The actual surface area does not count with the contact areas between the over imposed layers nor the contacting filaments. The actual surface area is corrected by estimating which is the percentage of air contacting surface per layer and then applying a correction factor to the surface area (multiplication or subtraction – See Table S4):

$$\textbf{Actual Surface area} \approx \textbf{Lenght}_{\textbf{Filam}} \times 2\pi\sqrt{\frac{a^2+b^2}{2}} \times \textbf{Correction factor} \text{ (Equation 5)}$$

The amount of filament per layer is similar. However, the percentage of contacting area varies, and the considerations made per each layer and model are shown in Table S4.

Further, the surface-to-volume ratio was calculated using the equation:

$$\textbf{Surface to volume ratio} \approx \frac{\textbf{Surface area}}{\textbf{V}_{\textbf{sliced model}}} \text{ (Equation 6)}$$

*2.8. Shape fidelity*

The shape fidelity of the height and diameter of the cookies after baking was quantified by:

$$\textbf{Fidelity} = \frac{\textbf{Dimension}_{\textbf{baked}} \cdot \textbf{100}}{\textbf{Dimension}_{\textbf{3Dmodel}}} \text{ (\%) (Equation 7)}$$

Cookies were analyzed in triplicate after baking for each sample.

*2.9. Colour*

A Minolta CR-310 portable colorimeter (Konica Minolta, Inc., Tokyo, Japan) measured the color of the doughs and 3D cookies. After calibration with a white standard, the values of $a^*$, $b^*$, and $L^*$ were measured.

The results were expressed as $\Delta E$ relative to the raw cookie dough as follow:

$$\boldsymbol{\Delta E^*} = [(\Delta \boldsymbol{L^*})^2 + (\Delta \boldsymbol{a^*})^2 + (\Delta \boldsymbol{b^*})^2]^{\frac{1}{2}} \text{ (Equation 8)}$$

*2.10. Total phenolic content and antioxidant activity*

Cookie extracts were obtained with ethanol 60 % in a 1:6 ratio (w:v) by ultrasonication for 30 min. After that, the samples were centrifuged for 10 min at 12,000 rpm in order to separate solid residues. Folin–Ciocalteu assay was based on a previously reported protocol (Ainsworth and Gillespie, 2007). All the reagents (Folin-Ciocalteu reagent, Na2CO3, Gallic acid, and DPPH) were purchased from Sigma-Aldrich (St. Louis, MO, USA). Briefly, the reaction mixture was prepared with 0.5 mL of extract, 0.05 mL of Folin-Ciocalteu reagent, and 0.5 mL of a 7 % aqueous $Na_2CO_3$ solution. The mixtures were incubated for 60 min at room temperature and the absorbance was measured at 760 nm in a microplate reader (Microplate Reader Synergy H1Hybrid Multi-Mode, BioTek, USA). A calibration curve of gallic acid was used to express the results as mgGAE/gdry product (milligrams of gallic acid equivalent per gram of dried sample).

DPPH assay was based on a reported procedure (Herald et al., 2012), with further modifications. Briefly, the absorbance of the ethanolic DPPH solutions was read at 515 nm and adjusted to 0.7 ± 0.02. Then, 25 µL of extract and 200 µL of DPPH solution were dispensed in a 96-well plate. A calibration curve of Trolox was used to express the results as µmolTR/gdry product (micromoles of Trolox equivalent per gram of dried sample). Each sample was measured in duplicate.

To express the results on a dry basis, the moisture contents were previously determined by oven-drying at 100 °C for 24 h using a Memmert Universal Oven UN110 (Memmert GmbH, Germany).

*2.11. Temperature sweep*

Oscillatory dynamic measurement and creep-recovery tests were performed in an HR-1 rheometer (TA Instruments, USA) equipped with a stainless steel parallel plate geometry (40 mm diameter, 1000 µm gap) within the linear viscoelasticity domain, in duplicate. Temperature-sweep profiles were measured in the range of 25 °C-180 °C at 5 °C/min and 1 Hz. The storage modulus, $G'$; loss modulus, $G''$; and loss factor, $\tan \delta$ (i.e., $G''/G'$) were recorded. All measurements were performed in duplicate using two different batches of dough.

*2.12. Design of experiments and Statistics*

In order to study the effect of the thickness (number of layers, Nº Layers) and porosity (infill %, Infill), baking temperature (Temp) and time (Time), on the properties of baked cookies, a full factorial design of these four independent variables was employed – Fig. 2 b. Fig. 2 clarifies the multiscale approach on which this study relies on and the range of Temp, Time, Nº Layer, and Infill analyzed. In addition to the convection acceleration effect, it was anticipated a protective effect of the encapsulation taking place on a smaller length scale, according to our previous results (Vieira et al., 2020). The dough fortified with encapsulated extract was 3D printed with 4 to 6 layers and Infill between 30 to 100 %, based on the 3D models presented in Fig. 2 c. The 3D printed dough was further baked for 10 to 20 min, at 140 to 180 °C. Moisture content, browning level, color variation ($\Delta E$), total phenolic content (TPC), antioxidant activity by 2,2-diphenyl-1-picrylhydrazyl-hydrate assay (DPPH), and (height and diameter) shape fidelity were the properties assessed.

Based on the water percentage in the baked 3D cookies, i.e., moisture, they were classified as overcooked (< 3 %), cooked (3-5 %), slightly cooked (5-10 %), or undercooked (> 10 %). The cookies also presented different levels of browning and were classified according to the following scores: - (no browning), + (slight browning), ++ (desired browning), and burnt (black and burnt smell). The data of the $2^4$ full-factorial design was modeled using Equation 9 to generate the mathematical models describing the impact of variables and their interactions on the properties mentioned above. Table S2 contains the experimental data, and Table S3 reports the equations, F-values, $R^2$ of the significant statistical models, i.e., moisture, DPPH, $\Delta E$, height, and diameter fidelity. The surface curves describing the main trends are in Fig. 3 a-d and will be further discussed.

The responses (Z) from the factorial design experiments were fitted to the following polynomial linear model:

$$Z = \beta_0 + \sum_{i=1}^{4}(\beta_i \cdot X_i) + \sum_{i=1}^{4}\left[\sum_{y=2}^{4}(\beta_{iy} \cdot X_i \cdot X_y)\right] + \sum_{i=1}^{4}\left[\sum_{y=2}^{4}\left[\sum_{w=3}^{4}(\beta_{iyw} \cdot X_i \cdot X_y \cdot X_w)\right]\right]$$

(Equation 9)

Where Z is the response (i.e., DPPH, moisture, diameter fidelity, height fidelity, total phenolic content (TPC), $\Delta E$) and $X_1$, $X_2$, $X_3$ and $X_4$ are the coded values of the independent variables (i.e., Time, Temp, Infill, and Nº Layers). $X_i \cdot X_y$, $X_i \cdot X_y \cdot X_y$ identify the interactions; $\beta_i$, $\beta_{iy}$, $\beta_{iyw}$ are the coefficients of the mentioned variables and the interactions; and $\beta_0$ is the independent term. Analysis of variance (ANOVA) was used to determine the significance and adequacy of the model using Statistica 10 (Tibco). Each condition was assessed with a triplicate of samples.

Independent two-sample Welch's t-test was used to compare the variance of two populations (cookies with and without pores, 4 layers, and baked at 180 °C for different times) assumed to not be the same. Each sample was composed of 3 specimens and the experiments were performed in triplicate.

## 3. Results and discussion

*3.1. Extraction and encapsulation of polyphenols from the grape skin*

Grape skin-derived extracts have numerous health benefits (Hogan et al., 2011; Hudson et al., 2007; Leifert and Abeywardena, 2008; Wu et al., 2018) associated with its consumption and are promising ingredients for functional foods. Herein, red grape skins extract rich in polyphenols was produced by ohmic heating-assisted extraction, following a previously reported protocol (Ferreira-Santos et al., 2019). This method enhances the extraction efficiency of high moisture compositions (El Darra et al., 2013; Lakkakula et al., 2004). Indeed, the antioxidant activity of the collected aqueous grape skins extract (1:10 solid/liquid ratio) increased by 5.5 fold when ohmic heating was applied –Table S1.

The extract was further incorporated into 6 % (w/v) maltodextrin and this mixture was nano spray-dried in a ratio of 1:1 of solids. The observation by scanning electron microscopy of the particle size and distribution analysis of the powders collected showed a typical round morphology and an average diameter of 1.6 ± 0.94 µm, containing 8 % below 500 nm - Fig. 1 a.

*3.2. Effect of the extract on the thermal viscoelastic properties*

The viscoelastic properties of three different doughs were monitored between 25 and 180 °C: blank dough, dough incorporated with free extract, and with the encapsulated extract. Assessing the thermal behavior of the dough informs whether the incorporation of the particles compromises its intrinsic heat-stability or the further shape stability during baking.

The temperature sweep curves as a function of elastic modulus, $G'$, and loss tangent, tan $\delta$, are presented in Fig. 1 b, c. $G'$ values relate to the energy stored and released by the dough, returning to its original form upon stress removal; therefore, reflecting the stiffness. $G''$ (not shown), or viscous modulus, reflects the energy dissipated with the flow. The ratio between the elastic and viscous moduli ($G''/G'$ or tan $\delta$) indicates which character predominates on the dough. The tan $\delta$ variations result from structure and state of the matter modifications by a series of reactions, i.e., protein denaturation, fat-melting, starch granules loss, and surface browning by the caramelization and Maillard reaction.

The fast increase of tan $\delta$ around 40 °C is related to butter melting. Between 80-90 °C, flour starch gelatinization(Kim et al., 2019), and thermal protein aggregation (Chevallier et al., 2002) (> 85 °C) occur, dropping the $G'$ and tan $\delta$. Above 90 °C, significant water evaporation initiates, increasing the dough stiffness. Around 120-140 °C, both $G'$ and tan $\delta$ decrease due to pyrolytic degradation and amylose leaching in the starch granules (Ahmed et al., 2013; Zhang et al., 2018). Meanwhile, the caramelization process also initiates (Purlis, 2010), especially above 130 °C.

Encapsulating the extract was required to avoid significant alteration on the thermal behavior of the dough. Whereas incorporating the encapsulated extract did not strongly impact the temperature sweep curves, slight shifts above 100 °C were visible in the dough with the free extract. Polyphenols react with proteins forming insoluble precipitates (Hagerman, 2012), which could affect the reactions.

Overall, the storage modulus was higher than the viscous modulus (tan δ < 1), which is characteristic of a solid-like dough. On top of that, the stabilized viscoelastic loss confirms the ability of this dough to ensure shape stability during baking not to be compromised (Kim et al., 2019). The dough contains xanthan gum, whose heat resistance capacity avoids further 3D structure collapse (Gimeno et al., 2004).

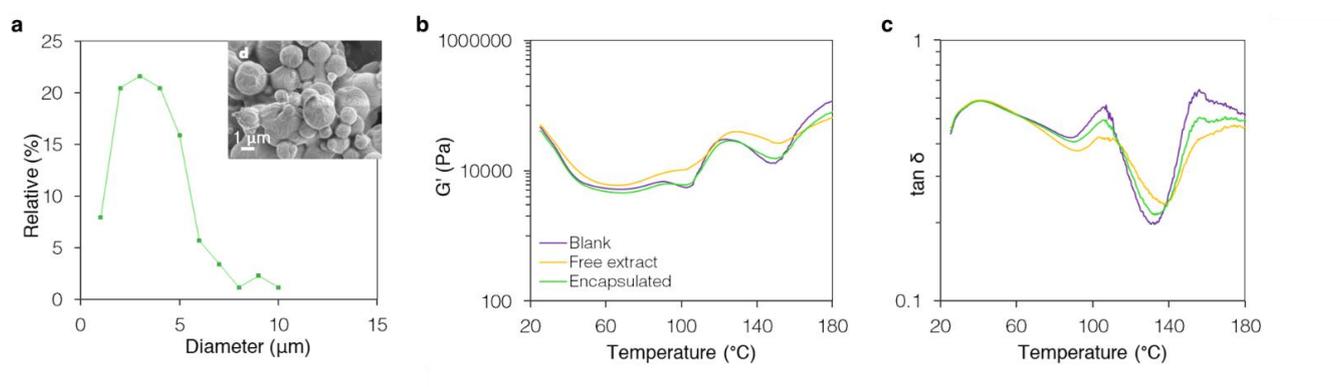

**Fig. 1.** Particle with encapsulated extract and rheological characterization of the doughs. a) Particle size distribution. b) Temperature (25-180 °C) sweep curves of the storage modulus of the blank dough (–), with enriched free extract dough (–) and encapsulated extract dough (–). c) Temperature sweep curves of the loss factor, i.e., tan δ ($G''/G'$), of the doughs during baking.

*3.3. Effect of extract encapsulation*

Per each 100 g of dough, 5 g of encapsulated extract yielded an increment of 1152 µmolTR/gdry of antioxidant activity and 36 mgGAE/gdry of phenolic compounds. However, a clear drop of such bioactivity was observed with increasing baking time and temperature, mostly in case the cookies browning and moisture had not significantly changed. In fact, the bioactivity of the free extract dropped by 64 % when it was incorporated in the non-encapsulated form on 5-layer cookies with 65 % infill, and baked for 15 min at 160 °C (i.e., central point). The antioxidant activity reduced to

421.3 µmolTR/gdry and the TPC to 29.5 mgGAE/gdry, also suggesting some degradation of phenolic compounds. Nevertheless, the encapsulation improved the antioxidant activity by 20 %. This effect of encapsulation corroborates our previous study on cookies fortified with encapsulated *A. platensis* derived extract (Vieira et al., 2020).

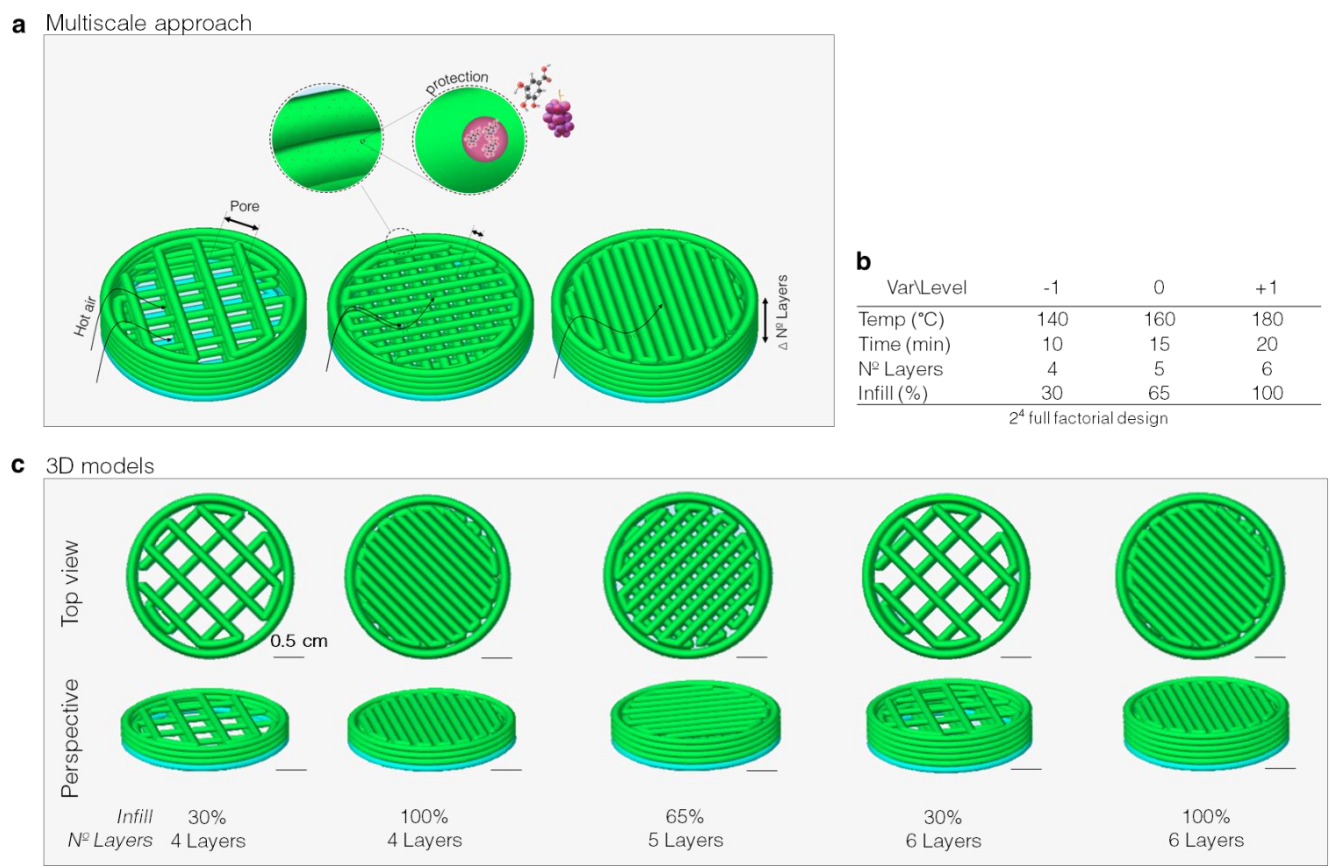

**Fig. 2.** 3D models and experimental design conditions tested on the 3D printed cookies. a) Multiscale strategy to model the functionality of cookies by encapsulating grape skin-derived polyphenols and by altering the surface-to-volume ratio and the porosity. b) Range of independent variables related to the cooking process and the design of 3D models assessed, i.e., temperature (Temp), baking time (Time), number of layers (Nº Layers), and infill percentage (Infill). c) 3D models used to print the cookie dough incorporated with encapsulated extract, with free extract and in the absence of extract, varying Nº Layers (4 to 6) and Infill (30 to 100 %).

*3.4. Effect of Time, Temperature, Nº Layers and Infill on moisture content*

The moisture model (Table S3) confirmed that all variables influenced the water content of the cookies. While, in general, increasing Temp and Time reduced the water content, augmenting the Nº Layers and Infill had the reverse effect – Fig. 3 a-b. The experimental data can be seen in Table 1.

According to the model, four Temp and Time combinations yielded intervals of Infill and Nº Layers where the cookies can be cooked: 160 °C for 20 min, 180 °C for 10 min, 180 °C for 15 min and 180 °C for 20 min – see Fig. S1. In specific, Fig. 3 b represents the moisture as a function of the Infill and Nº Layers, baked at 180 °C for 10 min. The model predicts that 4-layer cookies with 30 % infill can be cooked (i.e., moisture < 5%) at 180 °C in 10 min.

It is expected that reducing the Infill by introducing interconnected pores could alter the surface-to-volume ratio, facilitating the water removal, and accelerating the baking process (Zhang et al., 2018). The porosity and surface-to-volume ratio (Table S4) estimated for each of the sliced models was, respectively: 52.05 % and 1.08 $mm^{-1}$ (4 layers, 30 % Infill), 1.04 % and 1.14 $mm^{-1}$ (4 layers, 100 % Infill), 27.25 % and 0.62 $mm^{-1}$ (5 layers, 65 % Infill), 51.90 % and 0.81 $mm^{-1}$ (6 layers, 30 % Infill), and 2.35 % and 1.08 $mm^{-1}$ (6 layers, 100 % Infill). Increasing the Infill could reduce the surface-to-volume ratio and porosity. In contrast, more layers diminished the surface-to-volume ratio even though the porosities remained similar. However, the lowest ratio was estimated for 5 layers and 65 % of infill, which suggests a non-linear relation between Infill and Nº Layers.

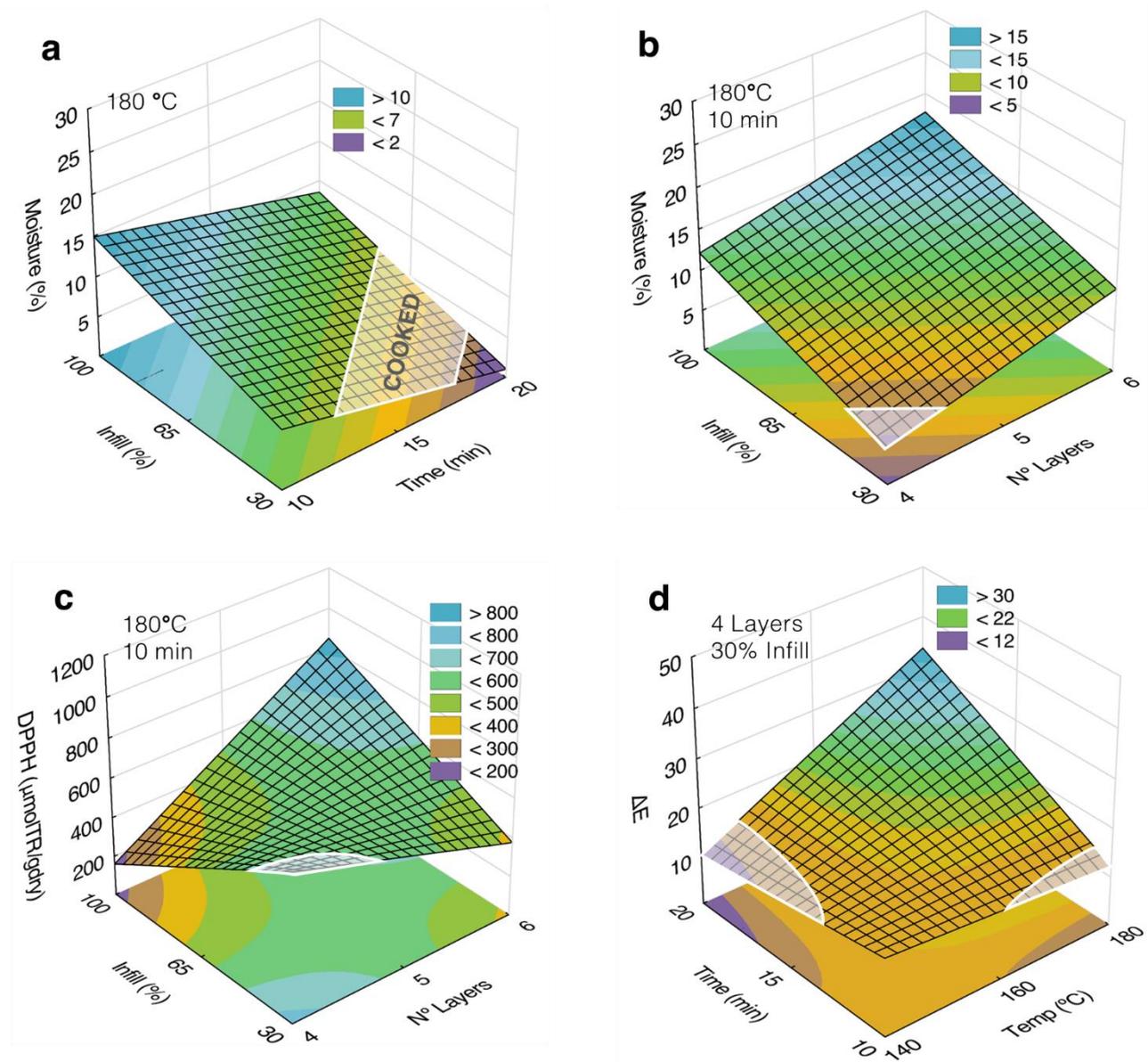

**Fig. 3.** Surface curves of the combined effect of cooking and design variables on features of the 3D printed cookies. a) Effect of the Infill and Time on the moisture when baking at 180 °C. b) Effect of the Infill and Nº Layers on the moisture when baking at 180 °C for 10 min. c) Effect of the Infill and Nº Layers on the antioxidant activity (DPPH) of cookies baked at 180 °C for 10 min. d) Effect of the Time and Temp on the 4-layer and 30 % infill cookies. Whether not indicated, the remaining variables were set on the central point (5 layers, 65 % Infill, 160 °C, and 15 min).

*3.5. Effect of Time, Temperature, Nº Layers and Infill on browning and antioxidant activity*

The DPPH radical scavenging activity model showed the Nº Layers and Infill as the main factors determining the antioxidant activity. Also, several other interactions with the Temp and Time were relevant factors in the equation – Table 1 and Table S3 for more details.

**Table 1.** Full factorial design, coded levels, and natural values of the 4 independent variables studied: oven temperature (Temp), cooking time (Time), number of layers (Nº Layers), and Infill (Infill %). Color, cooking level, TPC, and DPPH. Data are presented as average ± SD of 3 analyzed samples. Notes: * Blank 3D cookies; # 3D cookies with free extract.

|   | Temp (ºC) | Time (min) | Nº Layers | Infill (%) | ΔE | Browning (-, +, ++, burnt) | Moisture (%) | TPC (mgGAE/gdry) | DPPH (umolTR/gdry) |
|---|---|---|---|---|---|---|---|---|---|
|   | 140 | 20 | 6 | 100 | 9.88 | - | 28.65±5.30 | 371.53±14.05 | 375.58±52.46 |
|   | 160 | 15 | 5 | 65 | 8.98 | + | 14.61±1.86 | 438.39±4.4 | 410.05±86.86 |
|   | 180 | 10 | 6 | 100 | 10.82 | - | 21.05±0.76 | 466.70±14.24 | 933.63±95.40 |
|   | 180 | 10 | 4 | 30 | 14.03 | ++ | 9.31±1.61 | 436.74±20.45 | 754.89±41.89 |
|   | 180 | 20 | 4 | 100 | 31.15 | ++ | 4.57±0.70 | 221.37±9.40 | 584.46±110.59 |
|   | 160 | 15 | 5 | 65 | 12.36 | + | 15.63±1.27 | 234.28±21.74 | 413.96±37.30 |
|   | 140 | 20 | 4 | 30 | 10.75 | ++ | 6.28±1.50 | 368.33±62.37 | 384.80±61.58 |
|   | 180 | 20 | 6 | 30 | 37.78 | burnt | 2.34±0.69 | 396.86±72.18 | 382.87±20.8 |
|   | 140 | 10 | 6 | 30 | 14.02 | - | 20.20±0.93 | 668.81±38.21 | 388.29±23.31 |
|   | 140 | 10 | 4 | 100 | 15.82 | - | 17.23±4.92 | 522.25±25.97 | 447.75±28.30 |
|   | 140 | 20 | 4 | 100 | 11.89 | + | 13.47±0.94 | 364.38±17.18 | 621.62±60.28 |
|   | 180 | 20 | 4 | 30 | 45.33 | burnt | 1.08±0.07 | 325.49±25.65 | 446.87±33.34 |
|   | 140 | 10 | 6 | 100 | 19.58 | - | 24.32±0.8 | 491.30±12.57 | 936.68±149.28 |
|   | 160 | 15 | 5 | 65 | 13.00 | + | 8.50±0.56 | 230.68±16.09 | 403.56±33.37 |
|   | 180 | 10 | 6 | 30 | 15.68 | + | 7.94±0.39 | 278.62±19.68 | 438.90±47.12 |
|   | 140 | 20 | 6 | 30 | 14.00 | ++ | 8.45±0.49 | 469.77±25.99 | 887.44±25.12 |
|   | 180 | 10 | 4 | 100 | 14.17 | - | 11.06±0.54 | 324.06±29.71 | 303.52±10.43 |
|   | 180 | 20 | 6 | 100 | 26.01 | ++ | 6.98±0.66 | 369.43±8.58 | 770.72±100.24 |
|   | 160 | 15 | 5 | 65 | 12.96 | + | 8.19±0.58 | 275.65±6.38 | 457.52±29.21 |
|   | 140 | 10 | 4 | 30 | 20.64 | - | 11.80±6.05 | 249.58±15.61 | 316.96±24.61 |
| * | 160 | 15 | 5 | 65 | 13.34 | - | 9.57±0.13 | 62.35±2.47 | 242.82±29.03 |
| # | 160 | 15 | 5 | 65 | 9.58 | + | 8.43±1.61 | 159.8±10.19 | 349.68±8.24 |

The best interpretation is made in conjugation with moisture and browning results. Otherwise, some misleading conclusions can be taken. Some cases resulted in high DPPH, although corresponding to high moisture and not browned cookies. Overall, if the browning/Maillard reaction had not been initiated, nor a significant color change had been observed, the bioactivity was low (i.e., 300-400 µmolTR/gdry).

Fig. S1 shows the modeled DPPH contour graphs for all the combinations of Time and Temp as a function of the Infill and Nº Layers, highlighting where moisture was between 3 and 5 %. This analysis revealed that from a single dough formulation, the antioxidant activity of cooked cookies could be customized. The proper combination of the design and baking variables allowed to vary the bioactivity between 300 to 700 μmolTR/gdry. Among the regions with baked cookies, the highest DPPH values predicted by the model are obtained with 4-layer and 30 % infill cookies, baked for 10 min at 180 °C (~700 μmolTR/gdry) – Fig. 3 c. The experimental data of the DPPH (754 μmolTR/gdry) was in agreement with the one predicted by the model and accompanied by a high TPC (437 mgGAE/gdry). In other words, the bioactivity and the antioxidant content, respectively, increased by 115 % and 173 % when compared with the free extract cookies. The cookies with the second-best bioactivity, predicted by the models, would be printed with approximately 85-92.5 % infill, less than five layers, and baked for 20 min at 180 °C (~600 μmolTR/gdry). In the best scenario, the porosity and high temperature must have improved the heat transfer to the core, accelerating water removal, and the beginning of browning. Reducing the time might decrease the degradation of encapsulated compounds, meanwhile accelerating the production of new antioxidant species.

It should be noted that baking cookies in a shorter time usually do not raise concerns on texture alteration related to lack of time for starch granules gelatinization, such as the case of bread. In baked cookies, the percentage of starch granules that remain in their native structure is very high (Kulp et al., 1991). Furthermore, the full data suggested an additional point that might lead to high DPPH: baking the cookies 6-layer and 30 % infill at 140 °C for 20 min. However, those cookies were only slightly baked (5-10 % moisture).

A previous study reported that baking porous cereal foods at lower temperatures could preserve more the viability of incorporated probiotics (Zhang et al., 2018). The last noted condition might reflect the same behavior, but it seems baking over 20 min would be needed to reduce the moisture below 5 %. A further study would confirm this question and whether increasing the time would not compromise the high DPPH value observed. Interestingly, the same high levels of TPC and DPPH, and brown color were not observed on 4-layer cookies with 30 % infill baked in the same conditions. One plausible reason is that the higher surface-to-volume ratio of the cookies with 4-layers accelerated the baking such that the degradation of antioxidants was much more significant than the formation of new antioxidants.

The model and experimental data indicated that the best route to boost both the antioxidant activity and total phenolic content is producing 4-layer and 30 % infill cookies, which should be baked for 10 min at 180 °C. Besides, the experimental data called into attention that the proximity of 6-layers cookies with 30 % infill baked at 140 °C for 20 min might yield similar outputs; however, that would be outside of the predictive regions of the models.

*3.6. Progress of the properties of the 4-layer cookies with the baking time*

The porosity played a role in the properties of the cookies. Despite that, to better understand their evolution through baking, cookies were monitored at different times. Precisely, the DPPH radical scavenging activity, moisture, and phenolic content of the 4-layer cookies with and without pores were quantified after baking at 180 °C for 10, 15, and 20 min - Fig. 4 a-c.

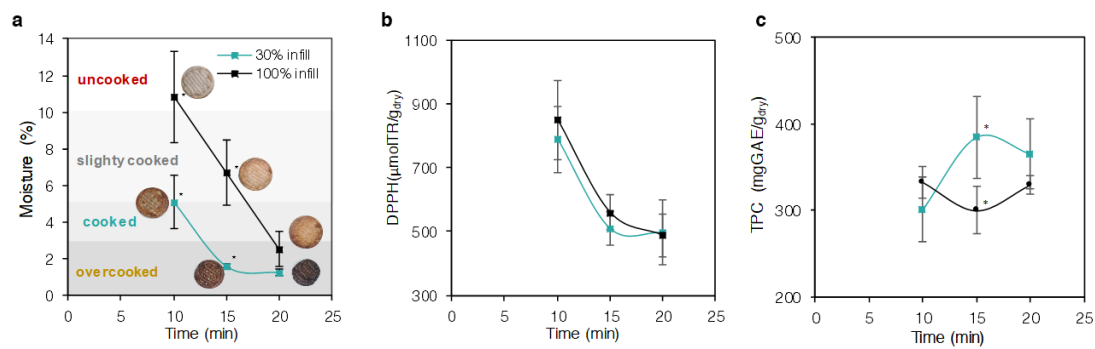

**Fig. 4.** Evolution of the properties of the 4-layer cookie with encapsulated extract with baking time, in the presence or absence of pores, i.e., 30 % (–) or 100 % (–) infill and baked at 180 °C. a) Moisture, b) DPPH,

and c) TPC vs. cooking time. All data are presented as average ± standard deviation (n = 3 and performed in triplicate). Significant differences between 30 % and 100 % infill cookies are indicated with * (p-value <0.05).

The cookies with 30 % infill were baked and browned in 10 min, although in the absence of porosity, approximately 20 min were required to reduce the moisture content below 5 % – Fig. 4 a. The pores did not influence the antioxidant activity, as shown in Fig. 4 b. Instead, the time appeared as the determining factor. The DPPH antioxidant activity exponentially dropped within 15 min and stabilized afterward. Although the TPC had been similar in 10 and 20 min, its tendency varied over the baking time - Fig. 4 c. In the first 15 min, there was a decay in the antioxidant content, whereas later on, a marginal increase was observed in the non-porous cookies (p-values: 0.028 - TPC 10 min *vs*. 15 min; 0.032 - TPC 15 min *vs*. 20 min; 0.032 - TPC 10 min *vs*. 20 min). In contrast, a significant increase that stabilized at 15 min was visible in the porous cookies (p-values: 0.006 - TPC 10 min *vs*. 15 min; 0.25 - TPC 15 min *vs*. 20 min; 0.013 - TPC 10 min *vs*. 20 min).

The formation of new antioxidant compounds during baking and simultaneous degradation elucidates why the DPPH values remained unchanged following the initial 15 min in both cases. The presence of interconnected pores in the cookies has indeed accelerated the formation of phenolic and other antioxidant compounds by at least 5 min earlier. Contrastingly, the antioxidants developed in the course of baking were insufficient to restore the initial DPPH radical scavenging activity loss when baking the cookies with no pores and at 180 °C.

*3.7. Shape Fidelity*

The analysis of the diameter and height fidelity of baked 3D printed cookies revealed that the cookies' thickness (i.e., height) increased by 18 % (6.72 mm) to 94 % (13.17 mm), and the diameter reduced up to 26 % (20.98 mm). The height was mostly determined by the Infill, and, to a lesser extent, by the Nº Layers. The more accurate thicknesses resulted from cookies with 30 % infill. The diameter was more consistent with the 3D models and relied on the Temp and the Infill. Baking at a higher Temp reduced the accuracy, although augmenting the Infill, slightly improved it - Fig. 5 a. Consequently, in the samples with 30 % infill, the cookies shrunk about 15 % - Fig. 5 b, c. The results suggest that water removal leads to structural shrinkage. Nevertheless, such imprecision is certainly amendable using a 3D model with a 15 % larger diameter.

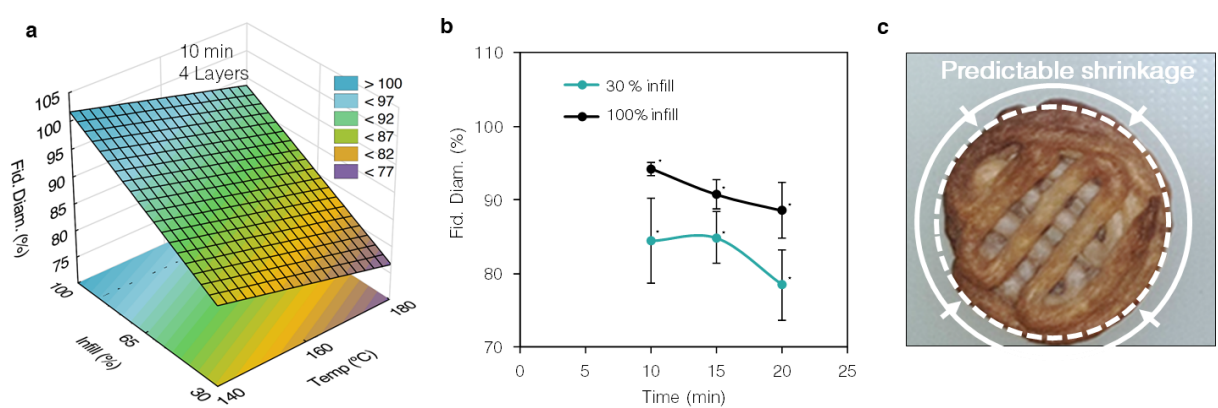

**Fig. 5.** Effect of Infill %, temperature, and oven time on diameter shrinkage. a) Response surface of the combined effect of the Infill and Temp on the diameter fidelity of the 4-layer cookies baked for 10 min. b) Shape fidelity variation as a function of the cooking time on the cookies with 30 % (–) and 100 % (–) infill and 4 layers baked at 180 °C. c) Actual diameter shrinkage on the baked cookies with the highest antioxidant activity (30 % infill, 4 layers, 180 °C, and 10 min baking). Data are presented as average ± standard deviation (n = 3 and performed in triplicate). Significant differences between 30 % and 100 % infill cookies are indicated with * (p-value <0.05).

Moreover, the shape fidelity indicates the cookie dough rises in the oven, even though it shrinks in the diameter direction. Also, the accuracy could be improved by introducing porosity. Producing cookies with pores accelerates moisture reduction and browning. The further brown crust, which is formed earlier than when the pores are absent, stabilizes the shape and diminishes the expansion.

**Conclusions**

Our results show how to employ 3D food printing to improve the antioxidant activity of cookies enriched with a grape skin extract. Encapsulating grape skin-derived extract prevents 20 % of the natural decay of the antioxidant activity during baking.

The experimental design pointed out that the best synergistic enhancement of both the antioxidant activity and phenolic content is achieved with 4-layer cookies and 30 % infill, baking for 10 min at 180 °C. In addition, the experimental data called into attention that conditions in the proximity of 6-layer cookies with 30 % infill, baked at 140 °C for 20 min, might yield similar outputs. In the first scenario, it was observed that introducing interconnected pores in the structure accelerates the moisture reduction and browning initiation. Consequently, the cookies were baked in a short time, new phenolic or other antioxidant molecules were generated earlier and, therefore, the oven time and related antioxidants degradation were reduced. In contrast, at low temperature, the degradation might be diminished, although a longer baking time to reduce the moisture below 5 % is needed.

The moisture and antioxidant activity model showed that a proper combination of the design and baking variables allows us to maximize the bioactivity of cooked cookies (moisture<5 %) to values between 300 to 700 µmolTR/gdry.

This represents a potential approach to enhance and customize the functional and healthy properties of cookies or other baked products.

**Author contributions**

This manuscript was written through the contributions of all authors. All authors have given approval to the final version of the manuscript.

**CRediT authorship contribution statement**

**All authors:** Writing - review & editing. **Sara M. Oliveira**, **Alice Gruppi**, and **Marta Vieira**: Writing - original drafts and experimental work. **Sara Oliveira**, **Alice Gruppi, Giorgia Spigno** and **Lorenzo Pastrana** – Conceptualization. **Sara M. Oliveira and Pablo Fuciños** – Data analysis.

**Declaration of competing interest**

The authors declare that they have no conflict of interest.

**Acknowledgments**

The research leading to these results has received funding from FODIAC – Food for Diabetes and Cognition, funded by European Union, under the call Marie Skłodowsk-Curie Research and Innovation Staff Exchange (Ref. H2020-MSCA-RISE-778388); PhD grantship from Fondazione di Piacenza e Vigevano (Doctoral School on the Agro-Food System, Università Cattolica del Sacro Cuore); Fondazione Cariplo through the project ReMarcForFood – Biotechnological strategies for the conversion of Winemaking by-products and their recycling into the food chain: development of new concepts of use, 2016-0740 grant.

**Supplementary data**

The following supplementary information is available

**References**

Ahmad, S.S., Morgan, M.T., Okos, M.R., 2001. Effects of microwave on the drying, checking and mechanical strength of baked biscuits. J. Food Eng. 50, 63–75. https://doi.org/10.1016/S0260-8774(00)00186-2

Ahmed, J., Almusallam, A.S., Al-Salman, F., AbdulRahman, M.H., Al-Salem, E., 2013. Rheological properties of water insoluble date fiber incorporated wheat flour dough. LWT-Food Sci. Technol. 51, 409–416.

Ainsworth, E.A., Gillespie, K.M., 2007. Estimation of total phenolic content and other oxidation substrates in plant tissues using Folin-Ciocalteu reagent. Nat. Protoc. 2, 875–877. https://doi.org/10.1038/nprot.2007.102

Akbari-Alavijeh, S., Shaddel, R., Jafari, S.M., 2020. Encapsulation of food bioactives and nutraceuticals by various chitosan-based nanocarriers. Food Hydrocoll. 105, 105774.

Ameur, L.A., Trystram, G., Birlouez-Aragon, I., 2006. Accumulation of 5-hydroxymethyl-2-furfural in cookies during the backing process: Validation of an extraction method. Food Chem. 98, 790–796.

Assadpour, E., Jafari, S.M., 2019. Advances in spray-drying encapsulation of food bioactive ingredients: From microcapsules to nanocapsules. Annu. Rev. Food Sci. Technol. 10, 103–131.

Augustin, M.A., Sanguansri, L., 2015. Challenges and Solutions to Incorporation of Nutraceuticals in Foods. Annu. Rev. Food Sci. Technol. 6, 463–477. https://doi.org/10.1146/annurev-food-022814-015507

Battino, M., Forbes-Hernández, T.Y., Gasparrini, M., Afrin, S., Cianciosi, D., Zhang, J., Manna, P.P., Reboredo-Rodr\'\iguez, P., Varela Lopez, A., Quiles, J.L., others, 2019. Relevance of functional foods in the Mediterranean diet: The role of olive oil, berries and honey in the prevention of cancer and cardiovascular diseases. Crit. Rev. Food Sci. Nutr. 59, 893–920.

Bigliardi, B., Galati, F., 2013. Innovation trends in the food industry: The case of functional foods. Trends Food Sci. Technol. 31, 118–129. https://doi.org/10.1016/j.tifs.2013.03.006

Charissou, A., Ait-Ameur, L., Birlouez-Aragon, I., 2007. Kinetics of formation of three indicators of the Maillard reaction in model cookies: Influence of baking temperature and type of sugar. J. Agric. Food Chem. 55, 4532–4539. https://doi.org/10.1021/jf063024j

Chevallier, S., Colonna, P., Buléon, A., Della Valle, G., 2000. Physicochemical behaviors of sugars, lipids, and gluten in short dough and biscuit. J. Agric. Food Chem. 48, 1322–1326. https://doi.org/10.1021/jf990435+

Chevallier, S., Della Valle, G., Colonna, P., Broyart, B., Trystram, G., 2002. Structural and chemical modifications of short dough during baking. J. Cereal Sci. 35, 1–10.

Chew, S., Tan, C., Pui, L., Chong, P., Gunasekaran, B., Nyam, K., 2019. Encapsulation technologies: A tool for functional foods development. Int. J. Innov. Technol. Explor. Eng. 8, 154–162.

Costa, C., Tsatsakis, A., Mamoulakis, C., Teodoro, M., Briguglio, G., Caruso, E., Tsoukalas, D., Margina, D., Dardiotis, E., Kouretas, D., Fenga, C., 2017. Current evidence on the effect of dietary polyphenols intake on chronic diseases. Food Chem. Toxicol. 110, 286–299. https://doi.org/10.1016/j.fct.2017.10.023

Daglia, M., 2012. Polyphenols as antimicrobial agents. Curr. Opin. Biotechnol. 23, 174–181. https://doi.org/10.1016/j.copbio.2011.08.007

Dankar, I., Haddarah, A., Omar, F.E.L., Sepulcre, F., Pujolà, M., 2018. 3D printing technology: The new era for food customization and elaboration. Trends Food Sci. Technol. 75, 231–242. https://doi.org/10.1016/j.tifs.2018.03.018

Day, L., Seymour, R.B., Pitts, K.F., Konczak, I., Lundin, L., 2009. Incorporation of functional ingredients into foods. Trends Food Sci. Technol. 20, 388–395. https://doi.org/10.1016/j.tifs.2008.05.002

De Rosso, V. V., Mercadante, A.Z., 2007. The high ascorbic acid content is the main cause of the low stability of anthocyanin extracts from acerola. Food Chem. 103, 935–943. https://doi.org/10.1016/j.foodchem.2006.09.047

Debicki-Pospisil, J., Lovrić, T., Trinajstić, N., Sabljić, A., 1983. Anthocyanin Degradation in the Presence of Furfural and 5-Hydroxymethylfurfural. J. Food Sci. 48, 411–416. https://doi.org/10.1111/j.1365-2621.1983.tb10754.x

Demirkol, E., Erdogdu, F., Palazoglu, T.K., 2006. Experimental determination of mass transfer coefficient: Moisture content and humidity ratio driving force approaches during baking. J. Food Process Eng. 29, 188–201. https://doi.org/10.1111/j.1745-4530.2006.00056.x

El Darra, N., Grimi, N., Vorobiev, E., Louka, N., Maroun, R., 2013. Extraction of Polyphenols from Red Grape Pomace Assisted by Pulsed Ohmic Heating. Food Bioprocess Technol. 6, 1281–1289. https://doi.org/10.1007/s11947-012-0869-7

Ferreira-Santos, P., Genisheva, Z., Pereira, R.N., Teixeira, J.A., Rocha, C.M.R., 2019. Moderate Electric Fields as a Potential Tool for Sustainable Recovery of Phenolic Compounds from Pinus pinaster Bark. ACS Sustain. Chem. Eng. 7, 8816–8826. https://doi.org/10.1021/acssuschemeng.9b00780

Gimeno, E., Moraru, C.I., Kokini, J.L., 2004. Effect of xanthan gum and CMC on the structure and texture of corn flour pellets expanded by microwave heating. Cereal Chem. 81, 100–107.

Godoi, F.C., Prakash, S., Bhandari, B.R., 2016. 3d printing technologies applied for food design: Status and prospects. J. Food Eng. 179, 44–54. https://doi.org/10.1016/j.jfoodeng.2016.01.025

Gueven, A., Hicsasmaz, Z., 2013. Pore structure in food: Simulation, measurement and applications, Pore Structure in Food: Simulation, Measurement and Applications. Springer.

https://doi.org/10.1007/978-1-4614-7354-1

Hagerman, A.E., 2012. Fifty years of polyphenol-protein complexes. Recent Adv. Polyphen. Res. 3, 71–97.

Herald, T.J., Gadgil, P., Tilley, M., 2012. High-throughput micro plate assays for screening flavonoid content and DPPH-scavenging activity in sorghum bran and flour. J. Sci. Food Agric. 92, 2326–2331. https://doi.org/10.1002/jsfa.5633

Hogan, S., Canning, C., Sun, S., Sun, X., Kadouh, H., Zhou, K., 2011. Dietary supplementation of grape skin extract improves glycemia and inflammation in diet-induced obese mice fed a Western high fat diet. J. Agric. Food Chem. 59, 3035–3041.

Hudson, T.S., Hartle, D.K., Hursting, S.D., Nunez, N.P., Wang, T.T.Y., Young, H.A., Arany, P., Green, J.E., 2007. Inhibition of prostate cancer growth by muscadine grape skin extract and resveratrol through distinct mechanisms. Cancer Res. 67, 8396–8405. https://doi.org/10.1158/0008-5472.CAN-06-4069

Karelakis, C., Zevgitis, P., Galanopoulos, K., Mattas, K., 2020. Consumer trends and attitudes to functional foods. J. Int. Food Agribus. Mark. 32, 266–294.

Kim, H.W., Lee, I.J., Park, S.M., Lee, J.H., Nguyen, M.H., Park, H.J., 2019. Effect of hydrocolloid addition on dimensional stability in post-processing of 3D printable cookie dough. LWT-Food Sci. Technol. 101, 69–75. https://doi.org/10.1016/j.lwt.2018.11.019

Kulp, K., Olewnik, M., Lorenz, K., Collins, F., 1991. Starch Functionality in Cookie Systems. Starch - Stärke 43, 53–57. https://doi.org/10.1002/star.19910430205

Kumar, Raman, Kumar, Ranvijay, others, 2020. 3D printing of food materials: A state of art review and future applications. Mater. Today Proc.

Lakkakula, N.R., Lima, M., Walker, T., 2004. Rice bran stabilization and rice bran oil extraction using ohmic heating. Bioresour. Technol. 92, 157–161. https://doi.org/10.1016/j.biortech.2003.08.010

Lavelli, V., Sri Harsha, P.S.C., Spigno, G., 2016. Modelling the stability of maltodextrin-encapsulated grape skin phenolics used as a new ingredient in apple puree. Food Chem. 209, 323–331. https://doi.org/10.1016/j.foodchem.2016.04.055

Le-Bail, A., Maniglia, B.C., Le-Bail, P., 2020. Recent advances and future perspective in additive manufacturing of foods based on 3D printing. Curr. Opin. Food Sci. 35, 54–64.

Leifert, W.R., Abeywardena, M.Y., 2008. Cardioprotective actions of grape polyphenols. Nutr. Res. 28, 729–737.

Lipton, J.I., Cutler, M., Nigl, F., Cohen, D., Lipson, H., 2015. Additive manufacturing for the food industry. Trends food Sci. Technol. 43, 114–123.

Mcclements, D.J., Decker, E.A., Park, Y., Weiss, J., 2009. Structural design principles for delivery of bioactive components in nutraceuticals and functional foods. Crit. Rev. Food Sci. Nutr. 49, 577–606. https://doi.org/10.1080/10408390902841529

Namiki, M., 1988. Chemistry of maillard reactions: Recent studies on the browning reaction mechanism and the development of antioxidants and mutagens, in: Advances in Food Research. Elsevier, pp. 115–184. https://doi.org/10.1016/S0065-2628(08)60287-6

Nayak, B., Liu, R.H., Tang, J., 2015. Effect of Processing on Phenolic Antioxidants of Fruits, Vegetables, and Grains - A Review. Crit. Rev. Food Sci. Nutr. 55, 887–918. https://doi.org/10.1080/10408398.2011.654142

Nikkhah, E., Khayamy, M., Heidari, R., Jamee, R., 2007. Effect of sugar treatment on stability of anthocyanin pigments in berries. J. Biol. Sci. 7, 1412–1417. https://doi.org/10.3923/jbs.2007.1412.1417

Palermo, M., Pellegrini, N., Fogliano, V., 2014. The effect of cooking on the phytochemical content of vegetables. J. Sci. Food Agric. 94, 1057–1070. https://doi.org/10.1002/jsfa.6478

Purlis, E., 2010. Browning development in bakery products - A review. J. Food Eng. 99, 239–249. https://doi.org/10.1016/j.jfoodeng.2010.03.008

Rodgers, S., 2016. Minimally Processed Functional Foods: Technological and Operational Pathways. J. Food Sci. 81, R2309–R2319. https://doi.org/10.1111/1750-3841.13422

Shtay, R., Keppler, J.K., Schrader, K., Schwarz, K., 2019. Encapsulation of (─)-epigallocatechin-3-gallate (EGCG) in solid lipid nanoparticles for food applications. J. Food Eng. 244, 91–100.

Souza, L. De, Madalena, D.A., Pinheiro, A.C., Teixeira, J.A., Vicente, A.A., Ramos, Ó.L., 2017. Micro- and nano bio-based delivery systems for food applications : In vitro behavior. Adv. Colloid Interface Sci. 243, 23–45.

Squillaro, T., Cimini, A., Peluso, G., Giordano, A., Melone, M.A.B., 2018. Nano-delivery systems for encapsulation of dietary polyphenols: An experimental approach for neurodegenerative diseases and brain tumors. Biochem. Pharmacol. 154, 303–317. https://doi.org/10.1016/j.bcp.2018.05.016

Thorvaldsson, K., Janestad, H., 1999. Model for simultaneous heat, water and vapour diffusion. J. Food Eng. 40, 167–172. https://doi.org/10.1016/S0260-8774(99)00052-7

Vieira da Silva, B., Barreira, J.C.M., Oliveira, M.B.P.P., 2016. Natural phytochemicals and probiotics as bioactive ingredients for functional foods: Extraction, biochemistry and protected-delivery technologies. Trends Food Sci. Technol. 50, 144–158. https://doi.org/10.1016/j.tifs.2015.12.007

Vieira, M. V, Oliveira, S.M., Amado, I.R., Fasolin, L.H., Vicente, A.A., Pastrana, L.M., Fuciños, P., 2020. 3D printed functional cookies fortified with Arthrospira platensis: Evaluation of its antioxidant potential and physical-chemical characterization. Food Hydrocoll. 105893.

Virág, D., Kiss, A., Forgó, P., Csutorás, C., Molnár, S., 2013. Study on Maillard-reaction driven transformations and increase of antioxidant activity in lysine fortified biscuits. Microchem. J. 107, 172–177. https://doi.org/10.1016/j.microc.2012.06.018

Wang, L., Boh, T., 2012. Health-Promoting Food Ingredients and Functional Food Processing, in: Nutrition, Well-Being and Health. pp. 201–229. https://doi.org/10.5772/25862

Wu, Z., Wu, A., Dong, J., Sigears, A., Lu, B., 2018. Skin extract improves muscle function and extends lifespan of a Drosophila model of Parkinson's disease through activation of mitophagy. Exp. Gerontol. 113, 10–17. https://doi.org/10.1016/j.exger.2018.09.014

Zhang, L., Lou, Y., Schutyser, M.A.I., 2018. 3D printing of cereal-based food structures containing probiotics. Food Struct. 18, 14–22. https://doi.org/10.1016/j.foostr.2018.10.002

**Supporting Information**

**How additive manufacturing can boost the bioactivity of baked functional foods**

Sara M. Oliveira[1], Alice Gruppi[1,2], Marta V. Vieira[1], Gabriela M. Souza[3], António A. Vicente[3], José A.C. Teixeira[3], Pablo Fuciños[1], Giorgia Spigno[2], Lorenzo M. Pastrana[1,*]

**Table S1.** Properties of the control doughs and grape skin extracts. Data are presented as average ± SD of 3 to 5 analyzed samples.

| Sample | Moisture (%) | TPC (mgGAE/$g_{dry}$) | DPPH (µmolTR/$g_{dry}$) |
|---|---|---|---|
| Freeze-dried Extract (1), 1:10 solid/liquid ratio | - | 48.90 ± 0.37 | 8907.94 ± 798.78 |
| Freeze-dried ohmic heated extract (2), 1:10 solid/liquid ratio | - | 18.94 ± 0.65 | 49261.16 ± 2384.12 |
| Encapsulated Extract | 0 | 33.62 ± 1.04 | 1145.26 ± 55.68 |
| Blank dough (160 °C, 15 min, 5 Layers) | 9.57 ± 0.11 | 62.35 ± 2.02 | 242.82 ± 23.70 |
| Baked (160 °C, 15 min, 5 Layers) dough with free ohmic heated extract | 8.43 ± 1.32 | 159.80 ± 8.32 | 349.68 ± 6.73 |

**Table S2.** Full factorial design, coded levels, and natural values of the 4 independent variables studied: oven temperature (Temp), cooking time (Time), number of layers (Nº Layers), and Infill (Infill %). Output results of the color, shape fidelity, cooking level, phenolic content, and antioxidant activity. Data are presented as average ± SD of 3 analyzed samples.

| Coded values | | | | Natural values | | | | Color | | | | Color | Shape Fidelity | | Cooking level | Antioxidants | |
|---|---|---|---|---|---|---|---|---|---|---|---|---|---|---|---|---|---|
| Temp | Time | Nº Layers | Infill | Temp (ºC) | Time (min) | Nº Layers | Infill (%) | L* | a* | b* | ΔE | Browning (-, +, ++, burnt) | Fid. Diam. (%) | Fid. Height (%) | Moisture (%) | TPC (mgGAE/gdry) | DPPH (µmolTR/gdry) |
| -1 | 1 | 1 | 1 | 140 | 20 | 6 | 100 | 60.46 | 4.12 | 11.87 | 9.88 | - | 105.63±1.14 | 164.73±3.58 | 28.65±5.30 | 371.53±14.05 | 375.58±52.46 |
| 0 | 0 | 0 | 0 | 160 | 15 | 5 | 65 | 62.81 | 5.20 | 14.24 | 8.98 | + | 99.20±2.03 | 145.24±4.92 | 14.61±1.86 | 438.39±4.4 | 410.05±86.86 |
| 1 | -1 | 1 | 1 | 180 | 10 | 6 | 100 | 59.34 | 4.60 | 10.96 | 10.82 | - | 97.09±1.49 | 163.49±2.19 | 21.05±0.76 | 466.70±14.24 | 933.63±95.40 |
| 1 | -1 | -1 | -1 | 180 | 10 | 4 | 30 | 55.92 | 4.05 | 9.60 | 14.03 | ++ | 78.20±1.46 | 135.79±3.57 | 9.31±1.61 | 436.74±20.45 | 754.89±41.89 |
| 1 | 1 | -1 | 1 | 180 | 20 | 4 | 100 | 47.85 | 15.00 | 27.62 | 31.15 | ++ | 86.61±2.16 | 193.15±5.4 | 4.57±0.70 | 221.37±9.40 | 584.46±110.59 |
| 0 | 0 | 0 | 0 | 160 | 15 | 5 | 65 | 60.36 | 6.40 | 16.26 | 12.36 | + | 91.12±2.23 | 156.25±2.01 | 15.63±1.27 | 234.28±21.74 | 413.96±37.30 |
| -1 | 1 | -1 | -1 | 140 | 20 | 4 | 30 | 64.17 | 6.83 | 17.43 | 10.75 | ++ | 89.46±0.98 | 136.83±3.51 | 6.28±1.50 | 368.33±62.37 | 384.80±61.58 |
| 1 | 1 | 1 | -1 | 180 | 20 | 6 | 30 | 36.10 | 14.47 | 21.59 | 37.78 | burnt | 74.19±1.44 | 118.55±6.40 | 2.34±0.69 | 396.86±72.18 | 382.87±20.8 |
| -1 | -1 | 1 | -1 | 140 | 10 | 6 | 30 | 60.09 | 8.62 | 17.50 | 14.02 | - | 79.25±0.24 | 113.24±4.25 | 20.20±0.93 | 668.81±38.21 | 388.29±23.31 |
| -1 | -1 | -1 | 1 | 140 | 10 | 4 | 100 | 54.12 | 3.93 | 8.58 | 15.82 | - | 98.19±2.17 | 191.37±5.24 | 17.23±4.92 | 522.25±25.97 | 447.75±28.30 |
| -1 | 1 | -1 | 1 | 140 | 20 | 4 | 100 | 58.49 | 7.09 | 12.14 | 11.89 | + | 97.54±1.53 | 190.03±5.02 | 13.47±0.94 | 364.38±17.18 | 621.62±60.28 |
| 1 | 1 | -1 | -1 | 180 | 20 | 4 | 30 | 23.80 | 8.30 | 8.22 | 45.33 | burnt | 75.69±1.10 | 148.21±4.61 | 1.08±0.07 | 325.49±25.65 | 446.87±33.34 |
| -1 | -1 | 1 | 1 | 140 | 10 | 6 | 100 | 50.02 | 5.89 | 6.79 | 19.58 | - | 106.63±1.34 | 171.43±2.20 | 24.32±0.8 | 491.30±12.57 | 936.68±149.28 |
| 0 | 0 | 0 | 0 | 160 | 15 | 5 | 65 | 59.64 | 7.99 | 16.40 | 13.00 | + | 85.76±0.23 | 157.50±2.75 | 8.50±0.56 | 230.68±16.09 | 403.56±33.37 |
| 1 | -1 | 1 | -1 | 180 | 10 | 6 | 30 | 55.59 | 8.40 | 15.52 | 15.68 | + | 80.04±1.20 | 121.92±2.85 | 7.94±0.39 | 278.62±19.68 | 438.90±47.12 |
| -1 | 1 | 1 | -1 | 140 | 20 | 6 | 30 | 56.83 | 7.65 | 13.99 | 14.00 | ++ | 83.97±1.42 | 131.50±1.67 | 8.45±0.49 | 469.77±25.99 | 887.44±25.12 |
| 1 | -1 | -1 | 1 | 180 | 10 | 4 | 100 | 56.90 | 8.02 | 14.62 | 14.17 | - | 90.89±0.12 | 193.82±1.49 | 11.06±0.54 | 324.06±29.71 | 303.52±10.43 |
| 1 | 1 | 1 | 1 | 180 | 20 | 6 | 100 | 54.48 | 13.14 | 29.06 | 26.01 | ++ | 97.31±0.59 | 194.05±3.76 | 6.98±0.66 | 369.43±8.58 | 770.72±100.24 |
| 0 | 0 | 0 | 0 | 160 | 15 | 5 | 65 | 59.67 | 8.19 | 16.43 | 12.96 | + | 82.35±1.09 | 148.57±2.64 | 8.19±0.58 | 275.65±6.38 | 457.52±29.21 |
| -1 | -1 | -1 | -1 | 140 | 10 | 4 | 30 | 48.73 | 7.45 | 9.68 | 20.64 | - | 83.93±0.06 | 140.10±9.55 | 11.80±6.05 | 249.58±15.61 | 316.96±24.61 |
| Blank 3D cookies | | | | 160 | 15 | 5 | 65 | 69.72 | -0.20 | 26.17 | 13.34 | - | 88.97±1.01 | 170.30±3.79 | 9.57±0.13 | 62.35±2.47 | 242.82±29.03 |
| 3D cookies w/ free extract (2) | | | | 160 | 15 | 5 | 65 | 60.54 | 8.37 | 14.96 | 9.58 | + | 87.77±0.09 | 159.52±3.68 | 8.43±1.61 | 159.8±10.19 | 349.68±8.24 |

**Table S3.** Equations of the models of the moisture content, Δ*E*, DPPH, diameter and height fidelity, $R^2$ and F-values (F table, $F_{tab}$; F calculated, $F_{calc}$ as the function of the factor and variables interactions with significant impact. The fitting for TPC was very poor, therefore it was not modeled.

| **Output** | **Equation** *(variables in coded value; alpha 5 %)* | $R^2$ | **F-value** (alpha= 5 %) |
|---|---|---|---|

| | | |
|---|---|---|
| Moisture (%) | 12.08300 - 4.08941x Temperature - 3.19292 x Time +2.81895 x Nº Layers + 3.74639 x Infill | $R^2$=.7743; Adj:.71411 | $F_{calc}$ (12.86) > $F_{tab}$ (3.06) |
| $\Delta E$ | 17.94547 + 4.90011 x Temperature + 3.87669 x Time + 6.81966 x Temperature x Time | $R^2$=.74869; Adj:.70157 | $F_{calc}$ (15.89) > $F_{tab}$ (3.24) |
| DPPH (µmolTR/gdry) | 533.004 +78.328 x Nº Layers +60.809 x Infill + 89.768 x Temperature x Time x Infill + 95.465 x Temperature x Nº Layers x Infill - 116.382 x Time x Nº Layers x Infill | $R^2$=.83788; Adj:.76305 | $F_{calc}$ (11.2) > $F_{tab}$ (2.92) |
| Fidelity Diam. (%) | 89.15217 - 4.03608 x Temp + 8.44731 x Infill (*independent variables in coded values*) | $R^2$=.78303; Adj:.75751 | $F_{calc}$ (30.67) > $F_{tab}$ (3.59) |
| Fidelity Height (%) | 155.7897 - 9.3998 x Nº Layers + 25.9952 x Infill | $R^2$=.91979; Adj:.91035 | $F_{calc}$ (97.47) > $F_{tab}$ (3.59) |

**Table S4.** Estimation of surface-to-volume ratio: 3D models highlighting the contact regions considered, the sliced model volume, the total length of extruded filament, the consideration made to determine the surface area correction factor, the actual surface area, and the surface-to-volume ratio.

| 3D model | Volume sliced model (mm$^3$) | Length Filam. (mm) | Surface area (mm$^2$) | Correction factor of surface area | Actual Surface area | Surface-to-volume ratio (mm$^{-1}$) |
|---|---|---|---|---|---|---|

| | | | | | | | |
|---|---|---|---|---|---|---|---|
| 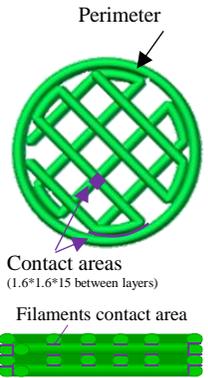 Perimeter / Contact areas (1.6*1.6*15 between layers) / Filaments contact area | 30 % Infill 4 Layers | 1285 | 3457 | 2387 | Layer 1 (L1) = 25 % <br> L2 = 25 % <br> L3 = 25 % <br> L4 = 25 % <br><br> ≈ 100 % and exclusion of contact areas: <br><br> - Contact between layers ≈ 1.6 mm*1.6 mm*15*3*2 surf ≈ 230.4 mm$^2$ <br><br> - Contact with perimeter ≈ [(6+7+8)*2*1.12*2]*4 layers*2 surfaces ≈ 752.6 mm$^2$ <br><br> - Contact between over-imposed perimeters ≈3*2 surfaces*π*[(27.6-26)/2]$^2$ = 12.06 mm$^2$ | 1394 | **1.08** |
| 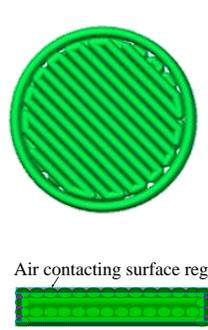 Air contacting surface region | 100 % Infill 4 Layers | 2652 | 7067 | 4879 | L1 ≈ 25 %*75 % <br> L2 ≈ 25 %*50 % <br> L3 = L2 <br> L4 = L1 <br><br> ≈ 62.5 % and exclusion of contact areas: <br><br> - Contact between over-imposed perimeters ≈3*2 surfaces*π*[(27.6-26)/2]$^2$ = 12.06 mm$^2$ | 3037 | **1.14** |

| 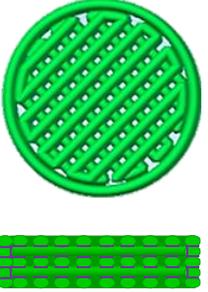 | 65 %<br>Infill<br>5 Layers | 2437 | 6567 | 4534 | L1 = 20 %<br>L2 = 20 %<br>L4 = L3 = L2<br>L5 = L1<br><br>- Contact between layers ≈ 1.6 mm*1.6 mm*72*4*2 surf ≈ 1474.6 mm$^2$<br><br>- Contact with perimeter ≈ [(5+5+4*6)*2*1.12*2]*5 layers*2 surfaces ≈ 1523.2 mm$^2$<br><br>- Contact between over-imposed perimeters ≈4*2 surfaces*π*[(27.6-26)/2]$^2$ = 16.08 mm$^2$ | 1520 | **0.62** |
| 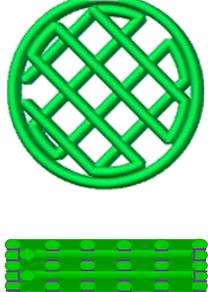 | 30 %<br>Infill<br>6 Layers | 1933 | 5200 | 3590 | L1 = 25 %<br>L2 = 25 %<br>L3 = 25 %<br>L4 = 25 %<br>L5 = 25 %<br>L6 = 25 %<br><br>≈ 100 % and exclusion of contact areas:<br><br>- Contact between layers ≈ 1.6 mm*1.6 mm*15*5*2 surfaces= 384 mm$^2$<br><br>- Contact with perimeter ≈ [(6+7+8)*2*1.6*2]*6 layers*2 surfaces = 1612.8 mm$^2$<br><br>- Contact between over-imposed perimeters ≈5*2 surfaces*π*[(27.6-26)/2]$^2$ = 20.11 mm$^2$ | 1573 | **0.81** |
| 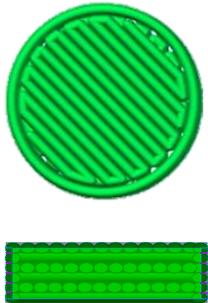 | 100 % Infill 6 Layers | 3925 | 10567 | 7296 | $L1 \approx 16.67\ \%*75\ \%$<br>$L2 = 16.67\ \%*50\ \%$<br>$L5=L4=L3=L2$<br>$L6 = L1$<br><br>$\approx 58.3\ \%$ and exclusion of contact areas:<br>- Contact between over-imposed perimeters<br>$\approx 5*2\ \text{surfaces}*\pi*[(27.6-26)/2]^2 = 20.11\ \text{mm}^2$ | 4236 | **1.08** |

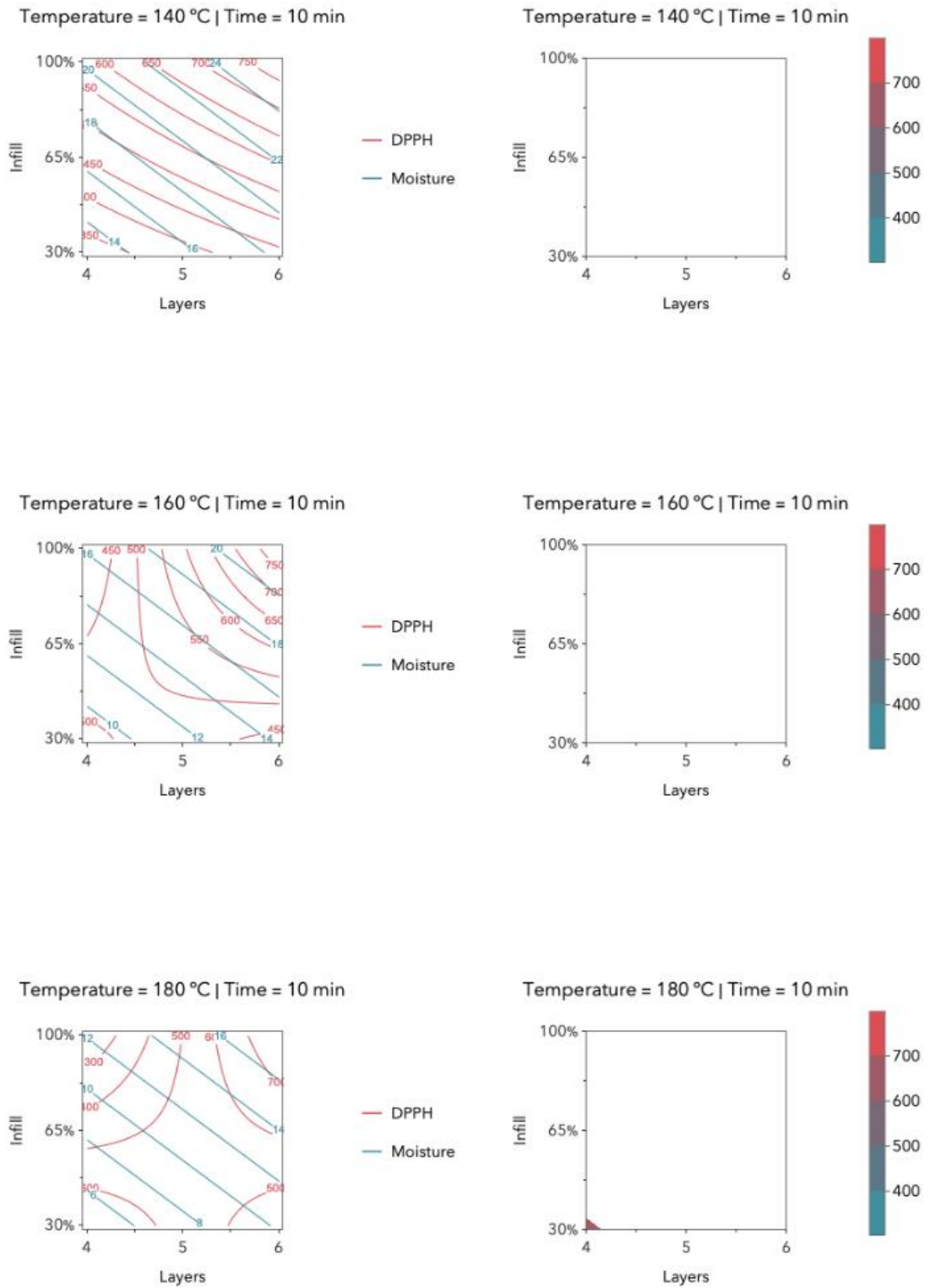

**Fig. S1, 1 of 3.** Contour graphs of the response surfaces of the moisture and DPPH models and the detection of baked cookies (moisture between 3-5 %).

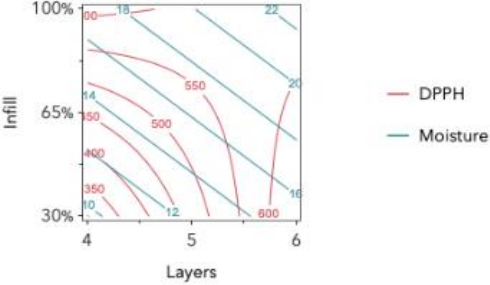 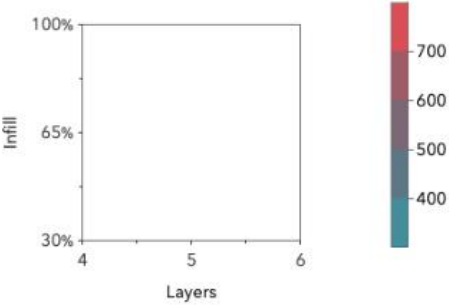

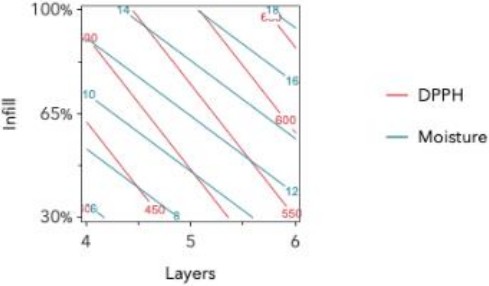 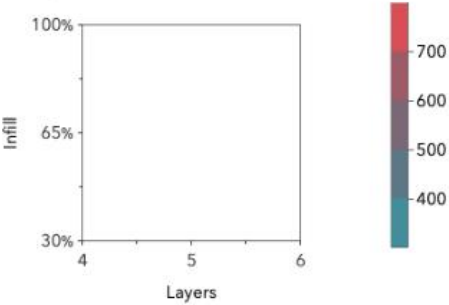

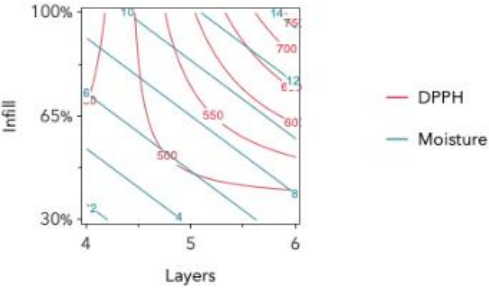 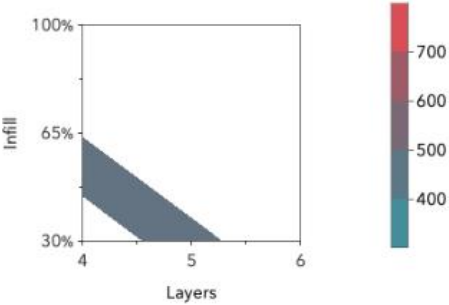

**Fig. S1, 2 of 3.** Contour graphs of the response surfaces of the moisture and DPPH models and the detection of baked cookies (moisture between 3-5 %).

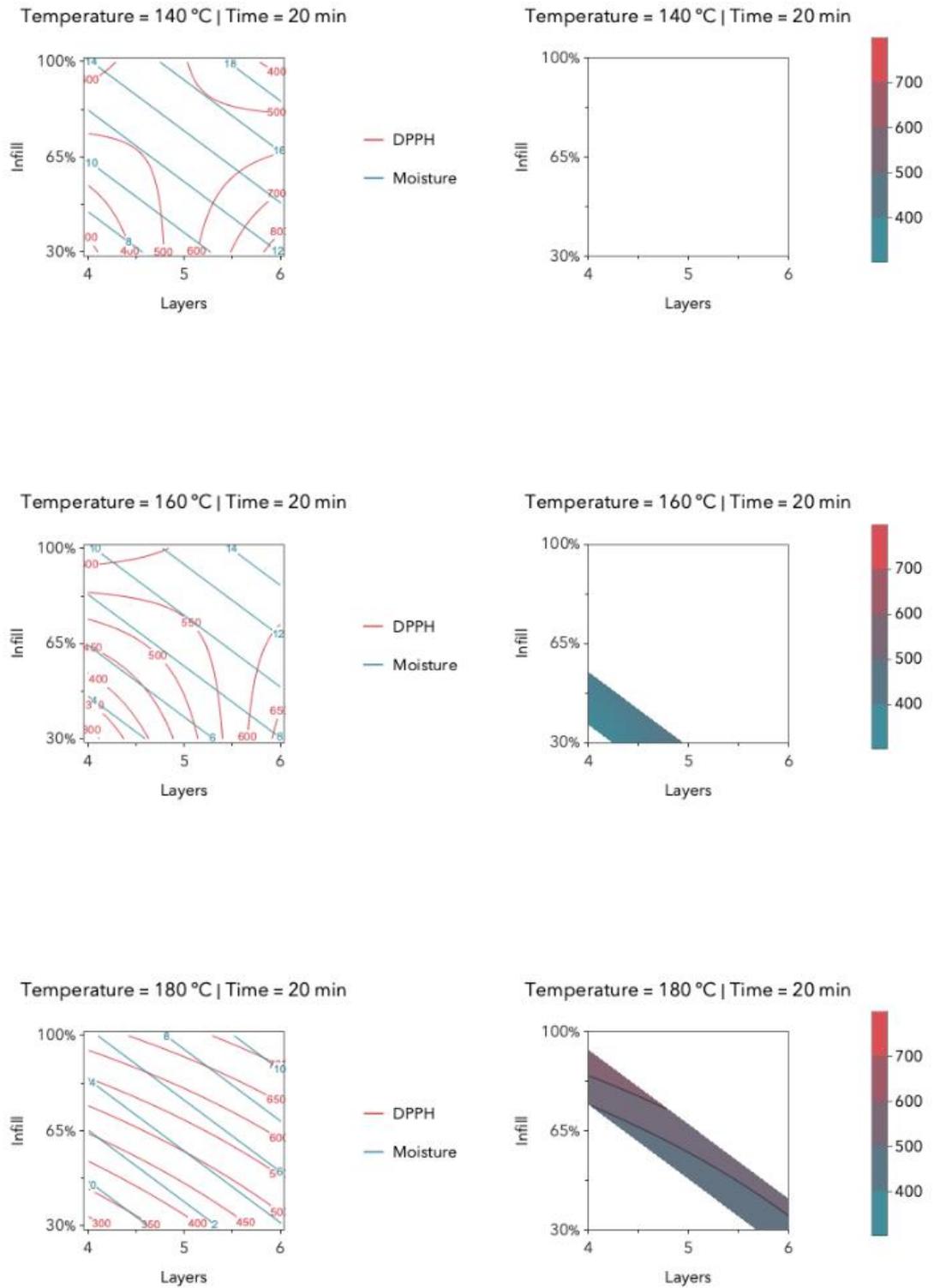

**Fig. S1, 3 of 3.** Contour graphs of the response surfaces of the moisture and DPPH models and the detection of baked cookies (moisture between 3-5 %).